\documentclass[onecolumn,preprint,superscriptaddress]{revtex4-1}

\usepackage[libertine,cmintegrals,cmbraces,vvarbb]{newtxmath}
\usepackage{amsmath}
\usepackage{graphicx}

\usepackage{bm}
\usepackage{float}
\floatstyle{plaintop} % places table captions above table
\restylefloat{table}
\usepackage[colorlinks=true, allcolors=blue]{hyperref}
\begin{document}

\title{Spin-0 bosons near rotating stars}

\author{E. O. Pinho}
\email{eduardoop3326@gmail.com}

\affiliation{Departamento de Física, CFM - Universidade Federal de \\ Santa Catarina; C.P. 476, CEP 88.040-900, Florianópolis, SC, Brazil}

\author{C. C. Barros Jr.}
\email{barros.celso@ufsc.br}

\affiliation{Departamento de Física, CFM - Universidade Federal de \\ Santa Catarina; C.P. 476, CEP 88.040-900, Florianópolis, SC, Brazil} 

\begin{abstract}
In this work we study the effects of rotating stars on the behavior of bosons close to their surfaces. For this task, metrics determined by the rotation of these stars will be taken into account. We will consider the Klein-Gordon equation in the Hartle-Thorne metric and in the one proposed by Berti et al. by considering some kinds of stars. Pions and Higgs bosons will be investigated as examples. By solving the equations, the main effect that may be observed is a rotational phase, $\delta_R$, that depends on the angular momentum of the star, $J$, in the wave function of the boson. Corrections of higher orders in $J$ are also investigated.
\end{abstract}

\maketitle

\section{Introduction}

The observation of the effects of the general theory of relativity
in physical systems has always been a challenging task. Since the first observation of light deflection by the Sun or the effect
of frame dragging due to the rotation of the earth \cite{schiff1960possible}, \cite{lense1918einfluss}, the
Lense-Thirring effect, which, after decades of experiments, was finally observed by the Gravity Probe B \cite{everitt2011gravity}, or even the recent detection of gravitational waves \cite{Abbott_2016},
we have witnessed great experimental efforts in order to obtain
reliable results.

A fundamental question in physics is how quantum mechanics and general relativity are related, or how general relativity affects quantum systems. At the moment a large number of 
such systems have been studied. Since the early studies from Parker \cite{Parker:1980hlc}, many others have been conducted, such as for example 
quantum oscillators \cite{Ahmed:2022tca}, \cite{Ahmed:2023blw},
\cite{Santos:2019izx}, \cite{Soares:2021uep}, \cite{Yang:2021zxo}, \cite{Rouabhia:2023tcl},
 magnetic fields in the Melvin metric \cite{Santos:2015esa},
 cosmic strings \cite{Santos:2016omw}, \cite{Santos:2017eef}, \cite{Vitoria:2018its},
the Casimir effect \cite{Bezerra_2016}, \cite{Santos:2018jba}
and many other kinds of systems \cite{Sedaghatnia:2019xqb}
\cite{Guvendi:2022uvz}, \cite{Vitoria:2018mun} \cite{Barros:2004ta}, 
\cite{Barros:2005jj}.

In this work, we will study the effect of the dragging of the spacetime
due to the rotation of stars in quantum systems, and for this
purpose, we will use spin-0 bosons as test particles.
A description of the spacetime of “slowly” rotating relativistic stars was constructed during the 1960s by James B. Hartle and Kip S. Thorne \cite{Hartle:1967he}, \cite{Hartle:1968si}, for both the exterior and interior regions of a star. The limits of the approximation, or in other words, the “slowness” of the stars, 
are in fact not so narrow when considering physical stars, such that this criterion is not
valid solely in the slow-rotation realm, yet it seems as if the solution is accurate for many systems, even for some that would not have been considered slow then.

Therefore, in this paper, we will use the external Hartle-Thorne metric as given by \cite{Hartle:1968si} and also the corrections proposed in \cite{berti2005}, and then write a
Klein-Gordon equation. We will solve the equations and then 
show numerical results for some typical systems.

This paper is divided into five sections beyond the introduction. In section \ref{section-2} we will describe the line elements we will be using in our calculations and in Sec. \ref{section-3} we will sketch the process by which we take the line element and expand it up to first order in the star's angular momentum $J$. Then, we will write the Klein-Gordon equation and solve it in various ways, in terms of several different special functions. In section \ref{section-4}, we will show an expansion that contains terms up to second order of the angular momentum $J$ and the resulting power series solution to the Klein-Gordon equation obtained in this spacetime. The resulting solution corresponds to the wave function
for a free particle plus additional mass and angular momentum corrections. We will also compare this solution with the one obtained in section \ref{section-3}, thereby showing that they have intrinsic similarities, despite having been obtained in conceptually different ways.
In Sec. \ref{section-5} the solution will then be studied for some physical systems
with different values of the mass and angular momentum and in
Sec. \ref{section-6} we will draw the conclusions of this work.

% Our first expansion, where we kept terms up to first order in the angular momentum $J$ of the star when substituted into the Klein-Gordon equation, became functionally equivalent to the same equation when using the Schwarzschild metric. However, we will show the process by which we arrived at an analytic—albeit power series—solution, corresponding to a well-known function, called the confluent Heun function, for the method we used to solve the equation may be useful in other contexts.

% In our second expansion, we kept terms up to second order in the angular momentum, and so, while it is more accurate, it is also much more complicated. By making several approximations—which were mostly reasonable since most numerical values in question are either very large or very small—we arrive at a series solution that corresponds to the free wave function, plus additional mass and angular momentum corrections. This approximation can be improved in several ways, however. The resulting solution is then studied and assessed as a solution in and of itself, and the mass and angular momentum contributions to the final solutions are measured for a number of physical systems.
%
\section{The Two Metrics}\label{section-2}
In this section, we will study the spacetime structure by means of metrics that take into account the rotation of stars. For this purpose we will consider the Hartle-Thorne metric \cite{Hartle:1967he}, \cite{Hartle:1968si} and the metric proposed by Berti et al. \cite{berti2005}. Since we are only interested in studying observable effects of bosons near rotating stars, we will only consider the solutions of the equations in an external region.

The Hartle-Thorne metric is a well-known spacetime metric that describes the geometry determined by slowly rotating stars. It was first obtained in a sequence of papers in the late 1960s and it has been used ever since in a variety of important works. The metric was initially constructed out of a very general canvas, that of an axially symmetric, stationary system that rotates uniformly and “slowly”. This “slow” rotation refers to the fluid's angular velocity $\Omega$, which must be slow enough for the  changes in pressure, energy density, and gravitational field due to rotation to be small, much less than unity. These considerations lead to the criterion presented by the authors which determines a scale factor
\begin{equation}
    \Omega^2 \ll \left( \frac{c}{R} \right)^2 \frac{G M}{R c^2}
    \equiv \frac{G M}{R^3}
\end{equation}
for a star of mass $M$ and radius $R$ that also means 
$R\Omega \ll c$, non relativistic particles inside the star.
We note, however, that this criterion is actually wide-reaching, and, as opposed to what may be thought at first glance, it does not necessarily correspond to a slow rotation when compared with the values of real stars \cite{Krastev_2008}.

A general line element may then be proposed in order to study a rotating object by considering it as an axially symmetric stationary system in terms of spherical coordinates
\begin{equation}\label{full-ht-metric}
    ds^2 = -H^2 dt^2 + E^2 dr^2 + r^2 K^2 [d\theta^2 + \sin^2{\theta} (d\phi - L dt)^2 ],
\end{equation}
%
% where $H$, $E$, $K$, and $L$ are functions of $r$ and $\theta$. The rotation is introduced as a perturbation and then the line element can be written as \cite{Hartle:1968si}
where $H$, $E$, $K$, and $L$ are functions of $r$ and $\theta$. Rotational effects are introduced as perturbations in these functions, and they become \cite{Hartle:1968si}
\begin{align}
    H^2 &= -\left( 1 - \frac{2M}{r} - \frac{2J^2}{r^4} \right) \Bigg\{ 1 + 2 \Bigg[ \frac{J^2}{M r^3} \left( 1 + \frac{M}{r} \right) + \frac{5}{8}\frac{Q - \frac{J^2}{M}}{M^3} {Q_2}^2 \left( \frac{r}{M} - 1 \right) \Bigg] P_2(\cos{\theta}) \Bigg\} \\
    E^2 &= \left( 1 - \frac{2M}{r} - \frac{2J^2}{r^4} \right)^{-1}  \Bigg\{ 1 - 2 \Bigg[ \frac{J^2}{M r^3} \left( 1 - \frac{5M}{r} \right) + \frac{5}{8}\frac{Q - \frac{J^2}{M}}{M^3} {Q_2}^2 \left( \frac{r}{M} - 1 \right) \Bigg] P_2(\cos{\theta}) \Bigg\} \\
    K^2 &= 1 + 2 \Bigg\{ - \frac{J^2}{M r^3} \left( 1 + \frac{M}{r} \right) + \\
    &\qquad + \frac{5}{8}\frac{Q - \frac{J^2}{M}}{M^3} \Bigg[ \frac{2M}{\sqrt{r(r-2M)}} {Q_2}^1\left( \frac{r}{M} - 1 \right) - {Q_2}^2\left( \frac{r}{M} - 1 \right) \Bigg] \Bigg\} P_2(\cos{\theta}) \\
    L &= \frac{2J}{r^3}
\end{align}
where $Q$ is the star’s mass quadrupole moment, $J$ its angular momentum, and
\begin{equation}
    \begin{split}
        {Q_2}^1(\zeta) &= \sqrt{\zeta^2 - 1} \left[ \frac{3\zeta^2 - 2}{\zeta^2 - 1} - \frac{3}{2}\zeta\ln{\frac{\zeta+1}{\zeta-1}} \right] \\
        {Q_2}^2(\zeta) &= - \frac{3\zeta^3 - 5\zeta}{\zeta^2-1} + \frac{3}{2}(\zeta^2-1)\ln{\frac{\zeta+1}{\zeta-1}} ,
    \end{split}
\end{equation}
are associated Legendre functions of the second kind. In \cite{abramowicz2003} it was pointed out that this metric is not consistently truncated, and for this reason, it presents some errors. A corrected metric was then proposed, but Berti et al. \cite{berti2005} studied that new metric and found some minor sign errors. They proposed the solution that will be
the second metric we will be using in this paper, which 
may be considered a corrected version of the original 
Hartle-Thorne metric. 
From \cite{berti2005}, the corresponding line element is given by
\begin{equation}\label{full-berti-metric}
    \begin{split}
        g_{rr} &= \left( 1 - \frac{2M}{r} \right)^{-1} [1 + j^2 G_1 + q F_2] + \mathcal{O}(\epsilon^3) \\
        g_{tt} &= -\left( 1 - \frac{2M}{r} \right) [1 + j^2 F_1 + q F_2] + \mathcal{O}(\epsilon^3) \\
        g_{\theta\theta} &= r^2 [1 + j^2 H_1 - q H_2] + \mathcal{O}(\epsilon^3) \\
        g_{\phi\phi} &= g_{\theta\theta} \sin^2{\theta} + \mathcal{O}(\epsilon^3) \\
        g_{t \phi} &= \left( \frac{2 j M^2}{r} \right) \sin^2{\theta} + \mathcal{O}(\epsilon^3)
    \end{split}
\end{equation}
where
\begin{equation}
    \begin{split}
        F_1 &= -p W + A_1 \\
        F_2 &= 5 r^3 p (3 u^2 - 1) (r - M) (2 M^2 + 6 M r - 3 r^2) - A_1  \\
        A_1 &= \frac{15 r (r - 2M) (1 - 3 u^2)}{16 M^2} \ln{\frac{r}{r - 2M}}  \\
        A_2 &= \frac{15 r (r - 2M^2) (3 u^2 - 1)}{16 M^2} \ln{\frac{r}{r - 2M}} \\
        G_1 &= p [(L - 72 M^5 r) - 3 u^2 (L - 56 M^5 r)] - A_1 \\
        H_1 &= A_2 + \frac{(1 - 3 u^2) (16 M^5 + 8 M^4 r - 10 M^2 r^3 + 15 M r^4 + 15 r^5)}{8 M r^4} \\
        H_2 &= - A_2 + \frac{5 (1 - 3u^2) (2M^2 - 3Mr - 3r^2)}{8 M r} \\
        L &= 80 M^6 + 8 M^4 r^2 + 10 M^3 r^3 + 20 M^2 r^4 - 45 M r^5 + 15 r^6 \\
        p &= \frac{1}{8 M r^4 (r - 2M)} \\
        W &= (r - M) (16 M^5 + 8 M^4 r - 10 M^2 r^3 - 30 M r^4 + 15 r^5)  \\
        &+ u^2(48 M^6 - 8 M^5 r - 24 M^4r^2 - 30 M^3 r^3 -60 M^2 r^4 + 135 M r^5 - 45 r^6) \\
        j &= \frac{J}{M^2} \\
        q &= \frac{Q}{M^3} \\
        u &= \cos{\theta}
    \end{split}
\end{equation}
and $\epsilon=\Omega /\Omega^*$ with $\Omega^*=\sqrt{M/R^3}$.
Observing these results, it seems reasonable to consider both metrics in the study of quantum systems, and since the corrections proposed in \cite{berti2005} are very small, we will check whether they are important in the behavior of spin-0 bosons near some kinds of stars.

\section{Klein-Gordon equation in the Hartle-Thorne metric}
\label{section-3}
In this section we will study the Klein-Gordon equation in the metrics presented in Sec. \ref{section-2} up to first order in the angular momentum ratio $j_1=J/r^2$. We must remark
that it is a suitable quantity for the analysis that we intend to do in this work since it presents small numerical values for the stars. For the sun, taking $r=R$ for example, $j_1=9.7\times 10^{-13}$; for a typical white dwarf, $j_1\sim 10^{-10}$; and for a typical neutron star,
$j_1\sim 10^{-5}$. As we can see, in a first approximation we may neglect terms of the order $j_1^2$ since they become numerically very small in the region outside of the star. In this section, we will work out the solutions up to first order in $j_1$, and in the next section, we will study the corrections up to $j_1^2$.  

In \cite{Hartle:1968si} the line element (\ref{full-ht-metric}) 
was expanded in powers of $M/r$, which, for the external region
of the Sun, for example, is smaller than $M/R\sim 2\times 10^{-6}$,
and when using numerical data relative to the Sun, 
an approximate form of this line element with an accuracy of the order of 1 part in $10^{15}$ has been proposed
\begin{equation}\label{expanded-ht-metric}
    \begin{split}
        ds^2 = &-\left[ 1 - \frac{2M}{r} + \frac{2Q}{r^3}P_2(\cos{\theta}) \right]dt^2 + \left[ 1 - \frac{2M}{r} + \frac{2Q}{r^3}P_2(\cos{\theta}) \right]^{-1}dr^2 \\
        & + r^2\left[ 1 - \frac{2Q}{r^3}P_2(\cos{\theta}) \right] \left[ d\theta^2 + \sin^2{\theta} \left( d\phi - \frac{2J}{r^3}dt \right)^2 \right].
    \end{split}
\end{equation}
We will study stars with nearly symmetric mass distributions
by taking $Q=0$, and since $J^2/r^6 \ll M/r$ for the external line element, we may write
\begin{equation}\label{final-ht-metric}
    \begin{split}
        ds^2 = &-\left( 1 - \frac{2M}{r} \right)dt^2 - \frac{4J}{r}\sin^2{\theta}dtd\phi \\
        &+ \left( 1 - \frac{2M}{r} \right)^{-1}dr^2 + r^2\left(d\theta^2 + \sin^2{\theta}d\phi^2 \right), 
    \end{split}
\end{equation}
which is the metric that will be considered in the calculations.

The Klein-Gordon equation for an arbitrary spacetime may be obtained by replacing the regular derivatives in the usual Klein-Gordon equation with covariant derivatives \cite{Birrell_N_D_1984}, \cite{Parker_Leonard2009-08-20}. In so doing, we arrive at the following:
\begin{equation}\label{kg-eq}
    \left[ -\frac{1}{\sqrt{-g}}\partial_\mu\big( g^{\mu\nu}\sqrt{-g}\partial_\nu \big) + \mu^2\right]\psi = 0.
\end{equation}
Because the components of our metric are not functions of $t$ or $\phi$, we can apply the method of separation of variables in the equation and obtain
\begin{equation}\label{wave-function-ansatz}
    \psi(t,r,\theta,\phi) = e^{i m\phi} e^{-i\omega t} \Theta(\theta) R(r),
\end{equation}
where $m=0,\pm 1, \pm 2 ...$, but we still need to solve the equations for $R(r)$ and $\Theta(\theta)$. The equation
for $\Theta (\theta)$ is
\begin{equation}
     \left[ \frac{d^2}{d\theta^2} + \frac{\cos\theta}{\sin\theta} \frac{d}{d\theta} + \lambda' - \frac{m^2}{\sin^2\theta} \right] \Theta (\theta) = 0,
\end{equation}
whose solution is simply the set of associated Legendre functions $P_l^m(\theta)$ with $\lambda'=l(l+1)$, where $l$ is an integer. The radial equation is
\begin{equation}\label{initial-form-radial-eq-ht-metric}
    \left( \mu^2 - \omega^2 A + \frac{l(l+1)}{r^2} + \frac{4 m \omega J A}{r^3}  \right) R - \frac{1}{r^2} \frac{d}{dr} \left( \frac{r^2}{A} \frac{dR}{dr} \right) = 0,
\end{equation}
where $A = \left( 1 - \frac{2M}{r} \right)^{-1}$. By substituting $R(r) = \frac{\sqrt{A}}{r} u(r)$, making the change of variables $x = \frac{r}{2M}$, and rearranging fractions, we arrive at the following equation
\begin{equation}\label{final-form-radial-eq-ht-metric}
    \frac{d^2u}{dx^2} + \left[ a + \frac{b}{x^2} + \frac{c}{x} + \frac{d}{(x-1)^2} + \frac{e}{x-1} \right]u = 0,
\end{equation}
where
\begin{equation}
    \begin{split}
        a &= 4M^2(\omega^2 - \mu^2) \\
        b &= \frac{1}{4} \\
        c &= \frac{1}{2} + l(l+1) - \frac{4m\omega J}{M} \\
        d &= \frac{1}{4} + 4M^2\omega^2 - \frac{4m\omega J}{M} \\
        e &= -\frac{1}{2} - 4M^2\mu^2 - l(l+1) + 8M^2\omega^2 + \frac{4m\omega J}{M}  \  .
    \end{split}
\end{equation}
This equation is the normal form of the confluent Heun equation, the canonical form of which is given by
\begin{equation}\label{canonical-heun-eq}
    \frac{d^2H}{dx^2} + \left( \alpha + \frac{\beta + 1}{x} +\frac{\gamma + 1}{x - 1} \right)\frac{dH}{dx} + \left( \frac{\sigma}{x} + \frac{\tau}{x-1} \right)H = 0,
\end{equation}
where $H(x) = HeunC(\alpha,\beta,\gamma,\delta,\eta,x)$ are the so-called confluent Heun functions with
\begin{equation}
    \begin{split}
        \sigma &= \frac{1}{2}(\alpha - \beta - \gamma -\alpha\beta - \beta\gamma) - \eta \\
        \tau &= \frac{1}{2}(\alpha + \beta + \gamma + \alpha\gamma + \beta\gamma) + \delta + \eta.
    \end{split}
\end{equation}
By means of an integrating factor, it is possible to relate the equations (\ref{final-form-radial-eq-ht-metric}) and (\ref{canonical-heun-eq}), and thus each of their coefficients,
\begin{equation}
    \begin{split}
        \alpha &= \pm 4M \sqrt{\mu^2-\omega^2} \\
        \beta &= 0 \\
        \gamma &= \pm 4\sqrt{\frac{m\omega J}{M} - M^2\omega^2} \\
        \delta &= 8M^2(2\omega^2 - \mu^2) \\
        \eta &= \frac{4m\omega J}{M} - l(l+1)  \ .
    \end{split}
\end{equation}
Having obtained these expressions, we have thus determined the exact analytical solution of the Klein-Gordon equation in the Hartle-Thorne metric given by eq. (\ref{final-ht-metric}).
It is possible to find a solution directly for eq. (\ref{canonical-heun-eq}) as a power series. We will use the asymptotic expansion for large $x$ found in \cite{Vieira_thesis_2018}, which was based, among others, on \cite{Ronveaux_A_1995}, a book specialized in Heun's differential equations. In the vicinity of the irregular singularity at infinity, the two solutions of equation (\ref{canonical-heun-eq}) can be written as
\begin{align}\label{eq:asymptotic-series-sol-heun}
    H_1(\alpha, \beta, \gamma, \delta, \eta, x) &\sim x^{-\left( \frac{\beta + \gamma + 2}{2} + \frac{\delta}{\alpha} \right)} \sum_{n=0}^\infty \frac{a_n(\alpha)}{x^n} \\
    H_2(\alpha, \beta, \gamma, \delta, \eta, x) &\sim x^{-\left( \frac{\beta + \gamma + 2}{2} - \frac{\delta}{\alpha} \right)} e^{-\alpha x} \sum_{n=0}^\infty \frac{a_n(-\alpha)}{x^n},
\end{align}
valid in the domain $-\pi - \textrm{arg}(\alpha + \eta) \le \textrm{arg}(x) \le \pi - \textrm{arg}(\alpha - \eta)$. The series coefficient $a_n$ is given by the following three-term recurrence relation:
\begin{equation}
    A_2 a_{n+2} + A_1 a_{n+1} + A_0 a_n = 0,
\end{equation}
where
\begin{align}
    A_2 &= -\alpha (n + 2) \\
    A_1 &= m(\alpha) + (n+1) \left(n + 2 + \alpha + \frac{2\delta}{\alpha} \right) \\
    A_0 &= \frac{\beta^2}{4} - \left( \frac{\gamma + 2}{2} + \frac{\delta}{\alpha} \right)^2 + n(n+1) - n\left( \frac{2\delta}{\alpha} + \gamma + 1 \right),
\end{align}
assuming $a_0 = 1$, and
\begin{align}
    &- \alpha a_1 + m(\alpha) a_0 = 0 \\
    &m(\alpha) = \eta - \frac{\gamma^2}{4} - \frac{\beta^2}{4} + \frac{a}{2} (\gamma + 1) + \frac{\delta}{\alpha} \left(\frac{\delta}{\alpha} + 1 + \alpha \right)
\end{align}
The negative power series given by (\ref{eq:asymptotic-series-sol-heun}) do not necessarily converge, and so we are required to truncate them at an appropriate spot. To choose said spot, we assume $a_{n+1}=0$, for some $n$, and then take $A_0=0$, which guarantees that, starting from a certain $n'$, $a_{n+2}$ will be 0 for any $n > n'$. Doing this procedure allows us to discover a relation that places a constraint on the coefficients of the confluent Heun equation (\ref{canonical-heun-eq}).
\begin{equation}\label{energy-spectrum}
    \frac{\beta^2}{4} - \left( \frac{\gamma + 2}{2} + \frac{\delta}{\alpha} \right)^2 + n(n+1) - n\left( \frac{2\delta}{\alpha} + \gamma + 1 \right) = 0
\end{equation}
By calling to mind how these coefficients relate to the physical quantities defined in the Hartle-Thorne metric (\ref{final-ht-metric}) and in the wave function ansatz (\ref{wave-function-ansatz}), it is clear that eq. (\ref{energy-spectrum}) defines an energy spectrum, $\omega(n)$, that may be solved numerically.

We may notice that some terms in eq. (\ref{initial-form-radial-eq-ht-metric}) are still very small, which makes it possible to obtain a simpler solution with the purpose of acquiring a better understanding of this system's physics by writing the solution in terms of more common special functions. With this purpose in mind, we can write eq. (\ref{final-form-radial-eq-ht-metric}) as
\begin{equation}\label{heun-eq-r-var}
    \frac{d^2u}{dr^2} + \left[ k^2 + \frac{a'}{r^2} + \frac{b'}{r} + \frac{c'}{(r-2M)^2} + \frac{d'}{r-2M} \right]u = 0,
\end{equation}
where
\begin{equation}\label{original-coeff-heun-eq-def}
    \begin{split}
        a' &= \frac{1}{4} \\
        b' &= \frac{1}{4M} + \frac{l(l+1)}{2M} - \frac{J m \omega}{M^2} \\
        c' &= \frac{1}{4} + 4M^2\omega^2 - \frac{2J m \omega}{M} \\
        d' &= -\frac{1}{4M} - 2M\mu^2 - \frac{l(l+1)}{2M} + 4M\omega^2 + \frac{J m \omega}{M^2} \\
        k^2 &= \omega^2 - \mu^2.
    \end{split}
\end{equation}
By observing that $a' /r^2$ and $b' /r$ are, numerically, several orders of magnitude smaller than
$c' / (r-2M)^2$ or $d' / (r-2M)$, for fixed parameters and $r$ values (outside the stars), a reasonable approximation to be made is to neglect these terms in eq. (\ref{heun-eq-r-var}).
If we also perform a change of variables $z = 2 i k (r - 2M)$, the equation becomes
\begin{equation}
    u''(z) + \left[ -\frac{1}{4} + \frac{1/4 - \lambda^2}{z^2} + \frac{\varepsilon}{z} \right] u(z) = 0,
\end{equation}
where $\varepsilon = d' / (2 i k)$ e $\lambda = \sqrt{1/4 - c'}$ and it is still relative to the line element (\ref{final-ht-metric}), but with the approximations shown previously.
The equation above is the so-called Whittaker equation, a modified version of the confluent hypergeometric equation, and it can be written in the form of Kummer's equation through a transformation of the function. This equation has two solutions, the Whittaker functions $M_{\varepsilon,\lambda}(z)$ and $W_{\varepsilon,\lambda}(z)$, which can be written, respectively, in terms of a Kummer function $M(A,B,z)$ and a Tricomi function $U(A,B,z)$, which are themselves solutions to the confluent hypergeometric equation \cite{abramo},
\begin{align}\label{whittaker-f}
    M_{\varepsilon,\lambda}(z) &= e^{- \frac{1}{2} z} z^{\frac{1}{2} + \lambda} M(\frac{1}{2} + \lambda - \varepsilon, 1 + 2\lambda, z), \\
    W_{\varepsilon,\lambda}(z) &= e^{- \frac{1}{2} z} z^{\frac{1}{2} + \lambda} U(\frac{1}{2} + \lambda - \varepsilon, 1 + 2\lambda, z).
\end{align}
Since the term proportional to $c'$ is still much smaller than the remaining terms of eq. (\ref{heun-eq-r-var}), and since in the region exterior to the stars $r \gg 2M$, we have
$r\simeq r-2m$, whereby the equation becomes, with good accuracy,
\begin{equation}\label{heun-eq-r-apr}
    \frac{d^2u}{dr^2} + \left[ k^2 + \frac{d'}{r} \right]u = 0.
\end{equation}
Supposing a solution with the form
\begin{equation}
u(r)= re^{ikr}w(r) 
\end{equation}
and a change of variables $z=2ikr$, we obtain the equation
\begin{equation}
    z w''(z) + (2 - z) w'(z) - \left( 1 - \frac{d'}{2 i k} \right) w(z) = 0,
\end{equation}
which has the solution
\begin{equation}
    w(z) = c_1 \, M\left( 1 - \frac{d'}{2ik}, 2, z \right) + c_2 \, U\left( 1 - \frac{d'}{2ik}, 2, z \right).
\end{equation}
Note that, for the physical problem we are studying, $c_2=0$. The Kummer function may be defined as 
\begin{equation}\label{def-kummer-f-power-series}
    M\left( 1 - \frac{d'}{2ik}, 2, z \right) = \sum_{s=0}^\infty \frac{\left( 1 - \frac{d'}{2ik} \right)_s}{(2)_s \, s!} z^s,
\end{equation}
which asymptotically behaves like a Bessel function. Thus, for large values of the variable $z$, which is precisely the region about which we are concerned, the solution may be written in the form
\begin{equation}\label{sol1}
    R(r) = \frac{e^{i\left[ kr + \frac{d'}{2k} \ln{r} \right]}}{r},
\end{equation}
or
\begin{equation}\label{soldeltaR}
R(r)=\Psi (r)e^{i\frac{jm\omega}{2k}\ln r} = \Psi (r)e^{i\delta_R},
\end{equation}
which explicitly shows the phase originated by the effect of the rotation of the star in the lowest order, $\delta_R$, which may be interpreted as a rotational phase. In the next section, we will investigate the solution for higher orders of $j_1$.

\section{Expansion of the Hartle-Thorne metric to second order in {\it J} }\label{section-4}
Become some mistakes were discovered in the Hartle-Thorne metric given by eq. (\ref{full-ht-metric}), since we are interested in studying terms of the order of $j_1^2$ and higher in the solution we have to use the Berti et al. metric
(\ref{full-berti-metric}) in our calculations. It is interesting to note that, 
if we keep the terms up to first order in $j_1$ and $Q=0$, eq. (\ref{expanded-ht-metric}) is recovered, which means that the results presented in the last section are correct and the differences occur for higher terms in $j_1$. So, in this section, we will derive an equation taking into account the terms that appear in the corrected metric and write down a solution based upon an infinite series ansatz. In order to observe the behavior of the corrections in the solution an easier way to do that is to solve the equations for fixed values of the angle $\theta$.
Taking these assumptions into account, the resulting radial equation becomes
\begin{equation}
    \left[ g^{rr} \partial_r^2 + \tilde{g} \partial_r \left( - \omega^2 g^{tt} - m^2 g^{\phi\phi} + m \omega g^{t \phi} - \mu^2 \right) \right] f(r) = 0,
\end{equation}
where
\begin{equation}
    \tilde{g} = \partial_r g^{rr} + \frac{1}{2g} g^{rr} \partial_r g.
\end{equation}
After some manipulations, the equation may be written in the form
\begin{equation}\label{final-berti-eq}
    \left( M_0 + \frac{M_1}{r} + \frac{M_2}{r^2} + \frac{M_3}{r^3} + \frac{M_4}{r^4} \right) g''(r) + \left( N_0 + \frac{N_1}{r} + \frac{N_2}{r^2} + \frac{N_3}{r^3} + \frac{N_4}{r^4} \right) g(r) = 0
\end{equation}
where
\begin{equation}\label{general-solution-with-integrating-factor}
    f(r) = \frac{A}{r - 2M} g(r)
\end{equation}
and
\begin{equation}
    \begin{split}
        M_0 &= 1 \\
        M_1 &= -6M \\
        M_2 &= 12 M^2 \\
        M_3 &= -8 M^3 \\
        M_4 &= 2 J^2 (5 - 12 u^2) \\
        N_0 &= \omega^2 - \mu^2 \\
        N_1 &= 4 M \mu^2 - 2 M \omega^2 \\
        N_2 &= m^2/(u^2 - 1) - 4 M^2 \mu^2 \\
        N_3 &= 4 m^2 M/(u^2 - 1) + 2 J m \omega + 2 M \\
        N_4 &= - 4 M^2 + 4 m^2 M^2/(u^2 - 1) - 4 J m M \omega + 2 J^2 \omega^2 (3 u^2 - 2).
    \end{split}
\end{equation}
This equation may be solved by the method presented in \cite{elizalde_1987}, where the solution has the form
\begin{align}
    g(r) &= \alpha \mathrm{e}^{\pm i F(r)} \label{g-def} \\
    F(r) &= k r + a_0 \ln{\frac{r}{2 M}} + \sum_{n = 1}^\infty a_n \left( \frac{2 M}{r} \right)^n. \label{F-def}
\end{align}
and the exponent $F(r)$ is a sum of functions and a power series of  $M/r$ where the $a_n$ coefficients must eventually be determined.
It is possible to find the first few coefficients by simply substituting this ansatz back into equation (\ref{final-berti-eq}), however, as $n$ increases the expressions for the coefficients become increasingly larger. The first five coefficients are given by
\begin{equation}\label{ansatz-coeff-sol}
    \begin{split}
        k &= \sqrt{\frac{N_0}{M_0}} \\
        a_0 &= \frac{-k^2 M_1 + N_1}{2 k M_0} \\
        a_1 &= \frac{1}{4 M k M_0} \Bigg( a_0 ^2 + 2 a_0 k M_1 + k^2 M_2 - N_2 \pm i a_0 M_0 \Bigg) \\
        a_2 &= \frac{1}{4 (2M)^2 k M_0} \Bigg( a_0^2 M_1 + 2 a_0 k M_2 + k^2 M_3 - N_3 - 4 M a_0 M_0 a_1 \\
        &- 4 M k M_1 a_1 \pm i a_0 M_1 \mp i 4 M M_0 a_1 \Bigg) \\
        a_3 &= \frac{1}{6 (2M)^3 k M_0} \Bigg( a_0^2 M_2 + 2 a_0 k M_3 + k^2 M_4 - N4 - 4 M a_0 M_1 a_1 \\
        &- 4 M k M^2 a_1 + 4 M^2 M_0 a_1^2 - 16 M^2 a_0 M_0 a_2 - 16 M^2 k M_1 a_2 + \\
        &\pm i a_0 M_2 \mp i 4 M M_1 a_1 \mp i 24 M^2 M_0 a_2 \Bigg).
    \end{split}
\end{equation}
For large values of $r$ we have
\begin{equation}
    R(r) \sim \frac{e^{i \left[ kr + a_0 \ln{r} \right]}}{r},
\end{equation}
which is the same result obtained in Sec. \ref{section-3}, in a very different way, shown in eq.(\ref{sol1}).

\section{Analyzing the solutions}\label{section-5}
In this section, we will investigate the solution (\ref{g-def}) for several physical configurations and observe how it changes depending on the mass, angular velocity, and energy of the star or of the particle whose motion the Klein-Gordon equation purports to describe. We will also study high-energy and low-energy test particles.
%
%\subsection{The solution to Berti's metric}
%
We will start by listing all physical configurations we chose in order to verify the size of the mass and angular momentum corrections. We selected three celestial bodies: the Sun \cite{Workman:2022ynf}, the neutron star PSR B1257+12 \cite{Zheng_2015},\cite{Wolszczan1992}, and the white dwarf PG 2131+066 \cite{Reed_2000}. All numerical values used in this section are given in Planck units, where $c=G=\hbar=1$. For the sun we have
\begin{equation}\label{sun-numbers}
    \begin{split}
        M_\odot &= 9.136 \times 10^{37} \\
        J_\odot &= 1.82 \times 10^{75} \\
        R_\odot &= 4.30835481 \times 10^{43},
    \end{split}
\end{equation}
for the neutron star,
\begin{equation}\label{ns-number}
    \begin{split}
        M_{ns} &= 1.4 M_\odot \\
        J_{ns} &= 0.00297 J_\odot \\
        R_{ns} &= 0.000015 R_\odot,
    \end{split}
\end{equation}
and for the white dwarf, the corresponding values are
\begin{equation}\label{wd-numbers}
    \begin{split}
        M_{wd} &= 0.608 M_\odot \\
        J_{wd} &= 0.1432 J_\odot \\
        R_{wd} &= 0.0186 R_\odot.
    \end{split}
\end{equation}
We also chose two test particles with very different masses, the pion with
mass $\mu_\pi$=0.139 GeV \cite{Workman:2022ynf}, and the Higgs boson with mass $\mu_{Higgs}$= 125.1 GeV,
which in Planck units are given by
\begin{equation}\label{particle-numbers}
    \begin{split}
        \mu_\pi &= 1.1056 \times 10^{-20} \\
        \mu_{Higgs} &= 1.0245 \times 10^{-17}
    \end{split}
\end{equation}
We separate eq. (\ref{F-def}) into four different parts: $\mathrm{A}$ is the real part of $F$ containing only mass terms; $\mathrm{B}$ is the imaginary part of $F$ containing only mass terms; $\Gamma$ is the real part of $F$ containing the angular momentum corrections, i.e., terms that are proportional to the angular momentum $J$ or to its square $J^2$; $\Delta$ is the imaginary part of $F$ containing the angular momentum contributions.
\begin{equation}\label{F-greek}
    F(r) = \mathrm{A}(r) + \mathrm{B}(r) + \Gamma(r) + \Delta(r)
\end{equation}
These functions have the same structure as $F(r)$, given by eq. 
(\ref{F-def}), and may be calculated by the sum of five terms
determined by different dependencies on the variable $r$. Together, as they should, they completely make up (\ref{F-def}). Then, by separating each type of contribution (mass and angular momentum) and also the real and imaginary parts of (\ref{F-def}), we can see that these newly defined functions are written as
\begin{equation}\label{ΑΒΓΔ-def}
    \begin{split}
        \mathrm{A}(r) &= kr + a_0 \ln{\frac{r}{2M}} + \alpha_1 \frac{2M}{r} + \alpha_2 \left( \frac{2M}{r} \right)^2 + \alpha_3 \left( \frac{2M}{r} \right)^3 \\
        \mathrm{B}(r) &= \beta_1 \frac{2M}{r} + \beta_2 \left( \frac{2M}{r} \right)^2 + \beta_3 \left( \frac{2M}{r} \right)^3 \\
        \Gamma(r) &= \gamma_2 \left( \frac{2M}{r} \right)^2 + \gamma_3 \left( \frac{2M}{r} \right)^3 \\
        \Delta(r) &= \delta_3 \left( \frac{2M}{r} \right)^3,
    \end{split}
\end{equation}
and the coefficients inside of it can be found by separating the coefficients (\ref{ansatz-coeff-sol}) in their real and imaginary parts and highlighting the ones with an angular momentum dependence, appearing in $\gamma_2$, $\gamma_3$, and $\delta_3$.

We will show the results for the six combinations of stars and particles given by eq. (\ref{sun-numbers})–(\ref{particle-numbers}) for particles with energies $\omega=2\mu$, $r=R$ (particles near the surface), $m=1$, and $\theta = \pi/4$.
These results are displayed in Tab. \ref{table-sun-pion}–\ref{table-white-higgs}.

In the tables, each column is related to one kind of dependence on the variable $r$. Each line in each table below corresponds to one of the four functions previously defined: $\mathrm{A(r)}$, $\mathrm{B(r)}$, $\Gamma(r)$ e $\Delta(r)$. Each column corresponds to the respective function's dependence on $r$, such that the first column corresponds to the contributions of $F(r)$ which are proportional to $r$, then to $\ln{r}$, and so on. For example, if we look at Tab. \ref{table-sun-pion}, the number $8.25031 \times 10^{23}$ found in the very first cell of the table, corresponds exactly to the term $kr$.
\begin{table}[H]
    \centering
    \resizebox{\columnwidth}{!}{%
    \begin{tabular}{|c|c|c|c|c|c|}
        \hline 
         & $r$ & $\ln{r}$ & $1/r$ & $1/r^2$ & $1/r^3$ \\
        \hline
        $\mathrm{A}(r=R)$ & $8.25031 \times 10^{23}$ & $5.04993 \times 10^{19}$ & $-1.71066 \times 10^{13}$ & $-3.6348 \times 10^{7}$ & $- 1.02727 \times 10^{2}$ \\
        \hline
        $\mathrm{B}(r=R)$ & 0 & 0 & $2.47395 \times 10^{-6}$ & $4.24684 \times 10^{-12}$ & $1.29492 \times 10^{-17}$ \\
        \hline
        $\Gamma(r=R)$ & 0 & 0 & 0 & $-5.66093 \times 10^{-13}$ & $-8.81302 \times 10^{-2}$ \\
        \hline
        $\Delta(r=R)$ & 0 & 0 & 0 & 0 & $6.86148 \times 10^{-37}$ \\
        \hline
    \end{tabular}%
    }
    \caption{Celestial object: Sun. Particle: Pion - contribution for each function according to the power of $r$.}
    \label{table-sun-pion}
\end{table}
\begin{table}[H]
    \centering
    \resizebox{\columnwidth}{!}{%
    \begin{tabular}{|c|c|c|c|c|c|}
        \hline 
         & $r$ & $\ln{r}$ & $1/r$ & $1/r^2$ & $1/r^3$ \\
        \hline
        $\mathrm{A}(r=R)$ & $1.23755 \times 10^{19}$ & $5.2965 \times 10^{18}$ & $-2.23527 \times 10^{18}$ & $-4.43284 \times 10^{17}$ & $-1.16929 \times 10^{17}$ \\
        \hline
        $\mathrm{B}(r=R)$ & 0 & 0 & $2.30902 \times 10^{-1}$ & $3.69947 \times 10^{-2}$ & $1.05281 \times 10^{-2}$ \\
        \hline
        $\Gamma(r=R)$ & 0 & 0 & 0 & $-7.48238 \times 10^{-6}$ & $-2.30951 \times 10^{8}$ \\
        \hline
        $\Delta(r=R)$ & 0 & 0 & 0 & 0 & $6.04614 \times 10^{-25}$ \\
        \hline
    \end{tabular}%
    }
    \caption{Celestial object: Neutron Star. Particle: Pion - contribution for each function according to the power of $r$.}
    \label{table-neutron-pion}
\end{table}
\begin{table}[H]
    \centering
    \resizebox{\columnwidth}{!}{%
    \begin{tabular}{|c|c|c|c|c|c|}
        \hline 
         & $r$ & $\ln{r}$ & $1/r$ & $1/r^2$ & $1/r^3$ \\
        \hline
        $\mathrm{A}(r=R)$ & $1.53456 \times 10^{22}$ & $2.2049 \times 10^{19}$ & $-3.39985 \times 10^{14}$ & $-2.36138 \times 10^{10}$ & $-2.18152 \times 10^{6}$ \\
        \hline
        $\mathrm{B}(r=R)$ & 0 & 0 & $8.0869 \times 10^{-5}$ & $4.53782 \times 10^{-9}$ & $4.52286 \times 10^{-13}$ \\
        \hline
        $\Gamma(r=R)$ & 0 & 0 & 0 & $-2.34386 \times 10^{-10}$ & $-2.81012 \times 10^{2}$ \\
        \hline
        $\Delta(r=R)$ & 0 & 0 & 0 & 0 & $1.52739 \times 10^{-32}$ \\
        \hline
    \end{tabular}%
    }
    \caption{Celestial object: White Dwarf. Particle: Pion - contribution for each function according to the power of $r$.}
    \label{table-white-pion}
\end{table}
\begin{table}[H]
    \centering
    \resizebox{\columnwidth}{!}{%
      \begin{tabular}{|c|c|c|c|c|c|}
        \hline 
         & $ r$ & $ \ln{r}$ & $ 1/r$ & $ 1/r^2$ & $ 1/r^3$ \\
        \hline
        $\mathrm{A}(r=R)$ & $7.64512 \times 10^{26}$ & $4.6795 \times 10^{22}$ & $-1.58518 \times 10^{16}$ & $-3.36818 \times 10^{10}$ & $-9.51912 \times 10^{4}$ \\
        \hline
        $\mathrm{B}$ & 0 & 0 & $ 2.47395 \times 10^{-6}$ & $4.24684 \times 10^{-12}$ & $1.29492 \times 10^{-17}$ \\
        \hline
        $\Gamma(r=R)$ & 0 & 0 & 0 & $-5.66093 \times 10^{-13}$ & $-8.16655 \times 10$ \\
        \hline
        $\Delta(r=R)$ & 0 & 0 & 0 & 0 & $7.40464 \times 10^{-40}$ \\
        \hline
    \end{tabular}  
    }
    \caption{Celestial object: Sun. Particle: Higgs boson - contribution for each function according to the power of $r$.}
    \label{table-sun-higgs}
\end{table}
\begin{table}[H]
    \centering
    \resizebox{\columnwidth}{!}{%
    \begin{tabular}{|c|c|c|c|c|c|}
        \hline 
         & $ r$ & $ \ln{r}$ & $ 1/r$ & $ 1/r^2$ & $ 1/r^3$ \\
        \hline
        $\mathrm{A}(r=R)$ & $1.14677 \times 10^{22}$ & $4.90798 \times 10^{21}$ & $-2.0713 \times 10^{21}$ & $-4.10768 \times 10^{20}$ & $-1.08352 \times 10^{20}$ \\
        \hline
        $\mathrm{B}(r=R)$ & 0 & 0 & $2.30902 \times 10^{-1}$ & $3.69947 \times 10^{-2}$ & $1.05281 \times 10^{-2}$ \\
        \hline
        $\Gamma(r=R)$ & 0 & 0 & 0 & $-7.48238 \times 10^{-6}$ & $-2.1401 \times 10^{11}$ \\
        \hline
        $\Delta(r=R)$ & 0 & 0 & 0 & 0 & $6.52476 \times 10^{-28}$ \\
        \hline
    \end{tabular}      
    }
    \caption{Celestial object: Neutron star. Particle: Higgs boson - contribution for each function according to the power of $r$.}
    \label{table-neutron-higgs}
\end{table}
\begin{table}[H]
    \centering
    \resizebox{\columnwidth}{!}{%
    \begin{tabular}{|c|c|c|c|c|c|}
        \hline 
         & $ r$ & $ \ln{r}$ & $ 1/r$ & $ 1/r^2$ & $ 1/r^3$ \\
        \hline
        $\mathrm{A}(r=R)$ & $1.42199 \times 10^{25}$ & 2.04316 $\times 10^{22}$ & $-3.15-45 \times 10^{17}$ & $-2.18816 \times 10^{13}$ & $-2.02149 \times 10^{9}$ \\
        \hline
        $\mathrm{B}(r=R)$ & 0 & 0 & $8.0869 \times 10^{-5}$ & $4.53782 \times 10^{-9}$ & $4.52286 \times 10^{-13}$ \\
        \hline
        $\Gamma(r=R)$ & 0 & 0 & 0 & $-2.34386 \times 10^{-10}$ & $-2.60399 \times 10^{2}$ \\
        \hline
        $\Delta(r=R)$ & 0 & 0 & 0 & 0 & $1.64829 \times 10^{-35}$ \\
        \hline
    \end{tabular}
    }
    \caption{Celestial object: White dwarf. Particle: Higgs boson - contribution for each function according to the power of $r$.}
    \label{table-white-higgs}
\end{table}

The tables above show the expected result that contributions diminish as the series goes and that the angular momentum contributions (corresponding to the functions $\Gamma(r)$ and $\Delta(r)$) are present, albeit very small. As we can see, the star for which the corrections are more significant is the neutron star. However, even the largest angular momentum corrections are small if compared with the other terms. For this reason, we will continue this analysis considering neutron stars.
By varying the physical quantities in question (such as the energy of the particle, and the angular momentum of the star) we will further explore our solution to the radial equation (\ref{final-berti-eq}).
We begin by defining three functions related to the solution which we will plot and which might give us some information on its status as a solution. These functions are
\begin{align}
    f_1(r) &= i F(r) \\
    f_2(r) &= \frac{e^{i F(r)}}{r - 2M} \\
    f_3(r) &= \frac{\Gamma(r) + \Delta(r)}{|F(r)|}
\end{align}
The function $f_1$ is precisely the exponent in the solution (\ref{g-def}). The function $f_2$ is eq. (\ref{g-def}), along with the integrating factor defined in (\ref{general-solution-with-integrating-factor}); the end result is that $f_2$ represents the radial solution of the wave function. The function $f_3$ is defined as the difference between the exponent (\ref{F-def}) and the mass contributions as defined in (\ref{F-greek}), divided by the absolute value of $F$. Since all that is left in the numerator of $f_3$ is the sum of angular momentum contributions $\Gamma(r) + \Delta(r)$ as defined in (\ref{F-greek}), this function tells us the approximate order of magnitude of the angular momentum contributions when compared to the mass contributions. 

\begin{figure}[H]
    \centering
    \includegraphics[width=\textwidth]{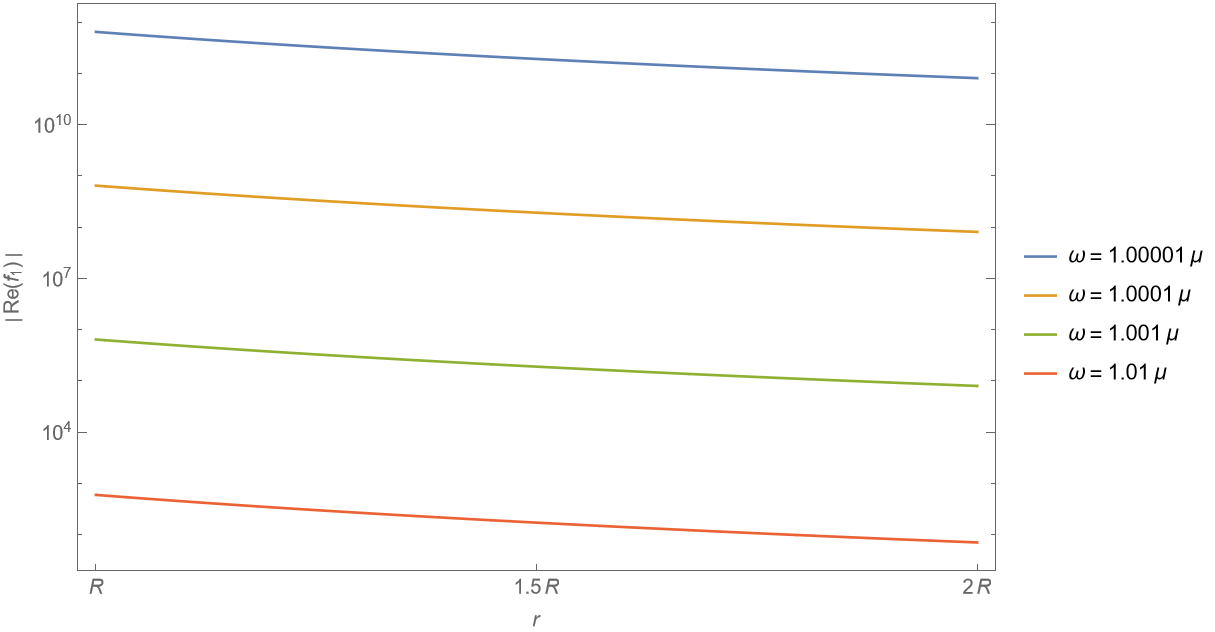}
    \caption{Absolute value of the real part of $f_1$ for some values of the energy. For even small changes to the energy, the value of $f_1$ changes substantially.}
    \label{graph-p1-vary-w-f-r-1}
\end{figure}

\begin{figure}[H]
    \centering
    \includegraphics[width=\textwidth]{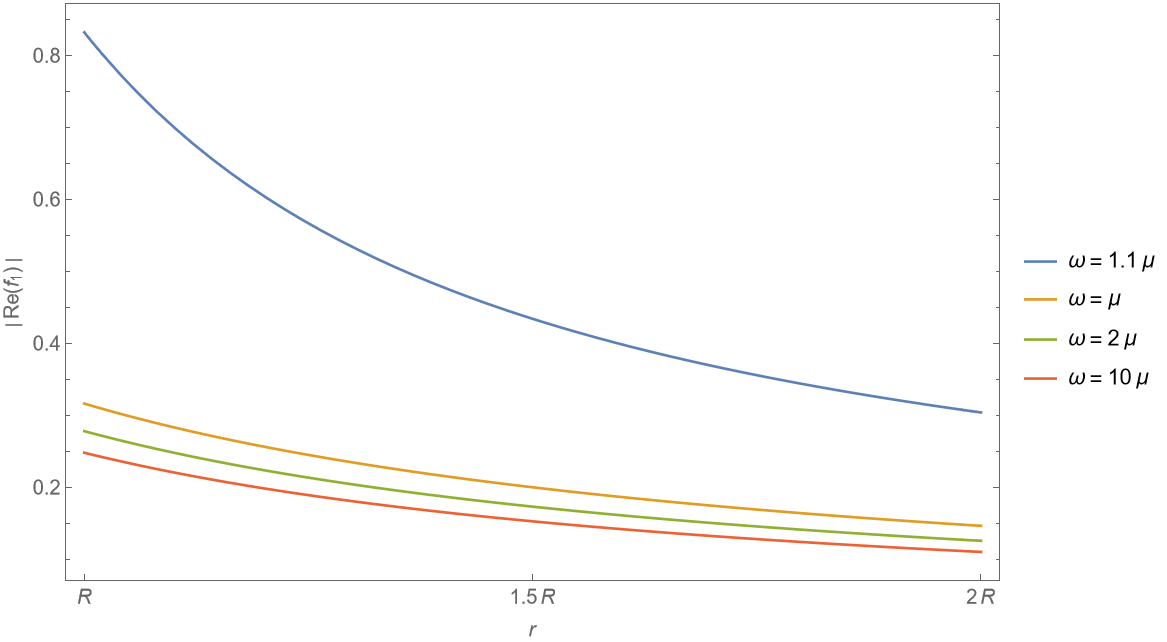}
    \caption{Absolute value of the real part of $f_1$ for higher
    values of the energy. It seems like for, high values of the energy, even large changes to its value do not change the function $f_1$ considerably.}
    \label{graph-p2-vary-w-f-r-2}
\end{figure}

Fig. 1 and Fig. 2 represent the function $f_1$ in two regions, in
Fig. 2 we show the results to the higher values of the energy.
As we can see, the major difference between figures \ref{graph-p1-vary-w-f-r-1} and \ref{graph-p2-vary-w-f-r-2} is that the value of the function stabilizes as the energy increases and even big differences in energy account for little changes in $f_1$. For lower energies, even with small changes, the value of $f_1$ changes considerably.

\begin{figure}[H]
    \centering
    \includegraphics[width=\textwidth]{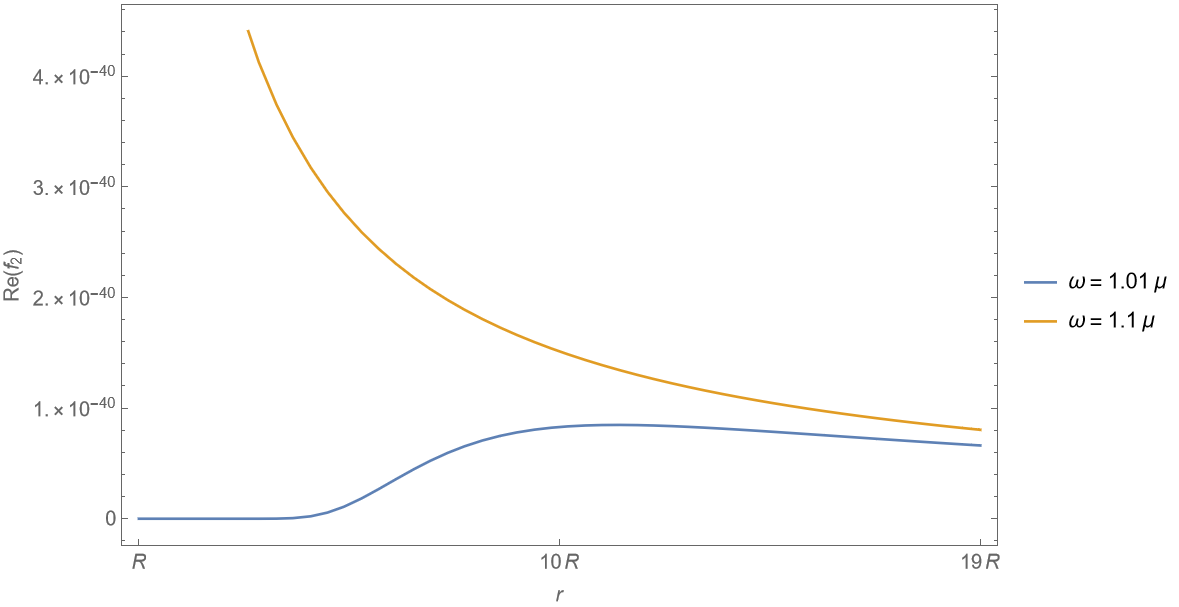}
    \caption{Real part of $f_2$ as function of $r$. Note that the effect for the lower energy may be a numerical oddity only. Nevertheless, the two curves start to coalesce for large $r$.}
    \label{graph-p9-exp-vary-w-f-r-1}
\end{figure}

\begin{figure}[H]
    \centering
    \includegraphics[width=\textwidth]{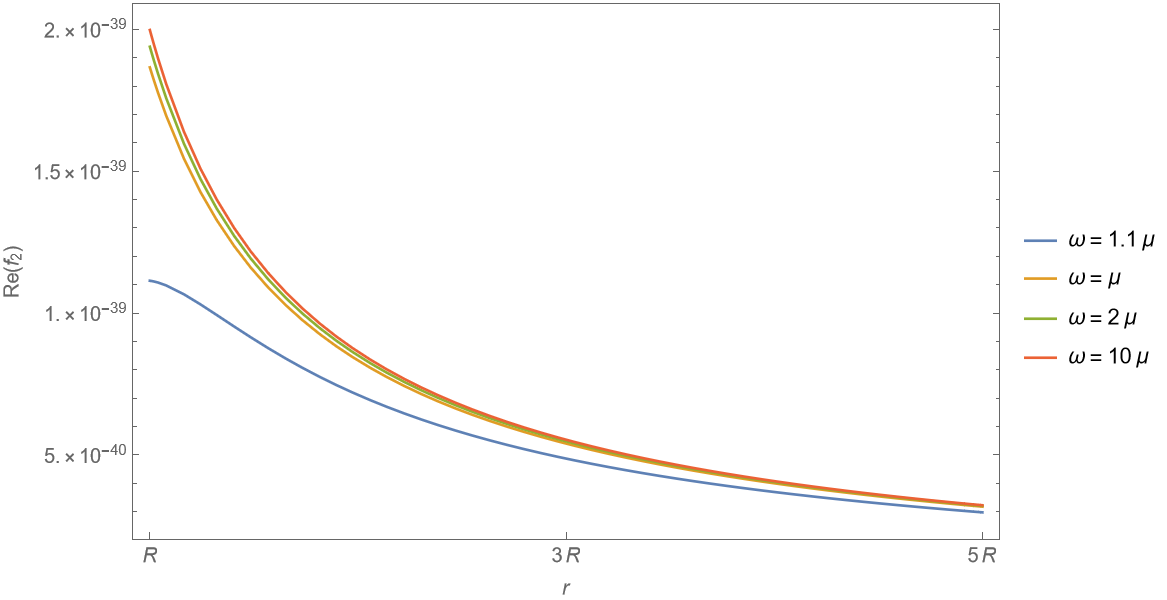}
    \caption{Real part of $f_2$ as function of $r$. The function $f_2$ seems to be independent of the energy for larger values of $r$.}
    \label{graph-p10-exp-vary-w-f-r-2}
\end{figure}

In Fig. 3 and 4 we represent the function $f_2$, the radial wave function, in two regions according to the values of the energies
of the pions. As we can see in Fig. \ref{graph-p9-exp-vary-w-f-r-1}, for large distances, far away from the surface of the star, the curves seem to coalesce, and this effect is especially visible for higher energies.

The last two graphs, Fig. 5 and 6, represent the function $f_3$, which shows the size of the angular momentum contribution as a variation in its value is done while maintaining the star’s original mass and radius fixed. We observe that the largest angular momentum selected, $J_6 = 10,000 J_{ns}$, where $J_{ns}$ is the angular momentum of the neutron star as given in (\ref{ns-number}), most likely cannot be reached for a real star as it exceeds the maximum possible value given in the literature \cite{Krastev_2008} for the angular velocity (and therefore angular momentum) of a neutron star. We used it regardless in order to show what happens to the solution for extreme rotations.

\begin{figure}[H]
    \centering
    \includegraphics[width=\textwidth]{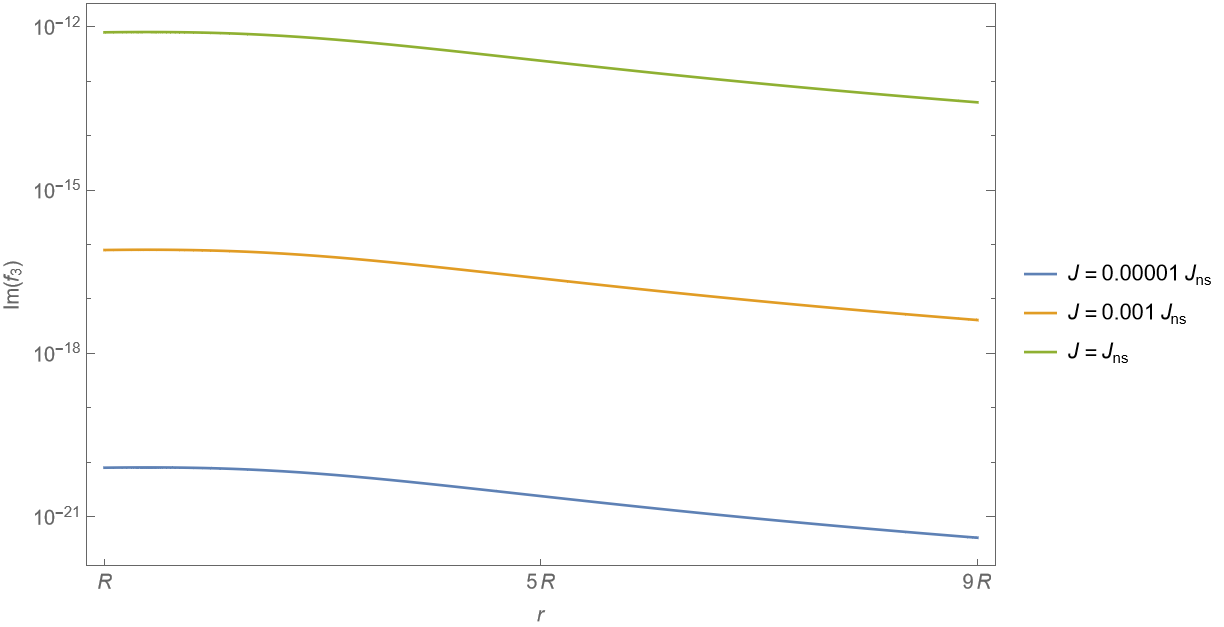}
    \caption{Imaginary part of $f_3$ as a function of $r$ for the actual angular momentum $J_{ns}$ of the neutron star and for variations of $J$. As previously stated, $f_3$ gives an estimate of the size of the angular momentum corrections, and this graph shows the shape of the curves is independent of the value of the angular momentum, but its value changes as expected.}
    \label{graph-p12-diff-vary-J-f-r-1}
\end{figure}

\begin{figure}[H]
    \centering
    \includegraphics[width=\textwidth]{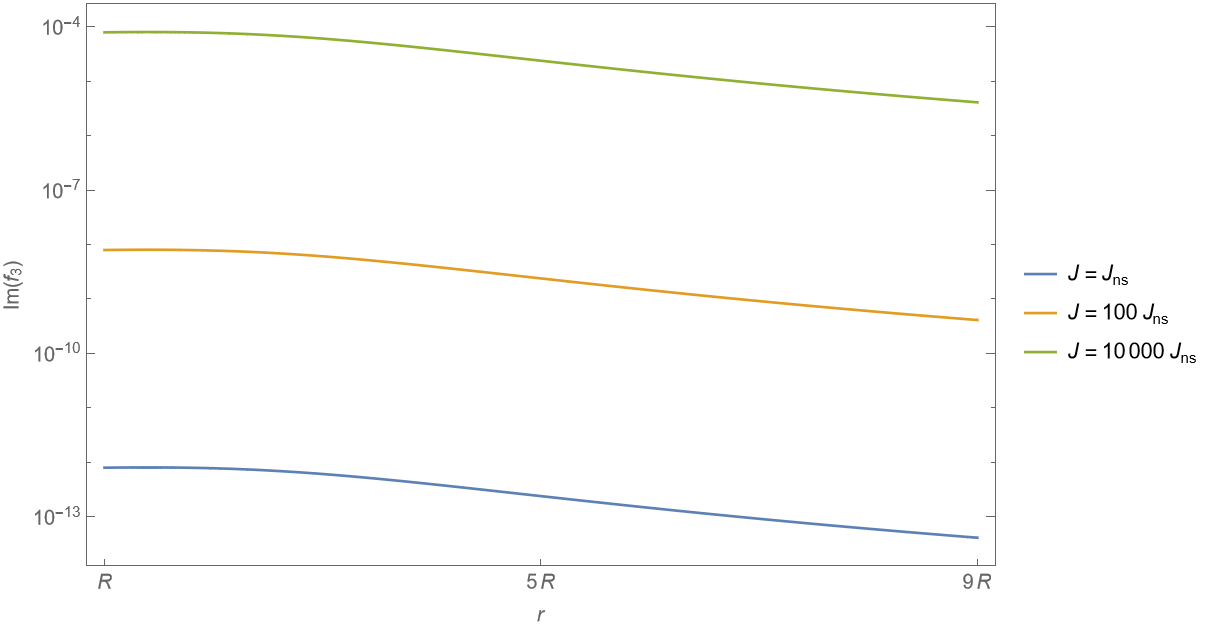}
    \caption{Imaginary part of $f_3$ as a function of $r$. The same as fig. \ref{graph-p12-diff-vary-J-f-r-1}, but for larger angular momenta.}
    \label{graph-p13-diff-vary-J-f-r-2}
\end{figure}

Fig. \ref{graph-p12-diff-vary-J-f-r-1} and \ref{graph-p13-diff-vary-J-f-r-2} show that the shape of the curves does not change considerably
when varying the angular momentum, however, the absolute value of the function does. This is the expected result: the larger the angular momentum of the neutron star, the larger will be its contribution to the solution. We can also investigate what happens to variations of the energy $\omega$ in the case where the angular momentum used was the neutron star’s own. The results are shown in Fig. 7.

\begin{figure}[H]
    \centering
    \includegraphics[width=\textwidth]{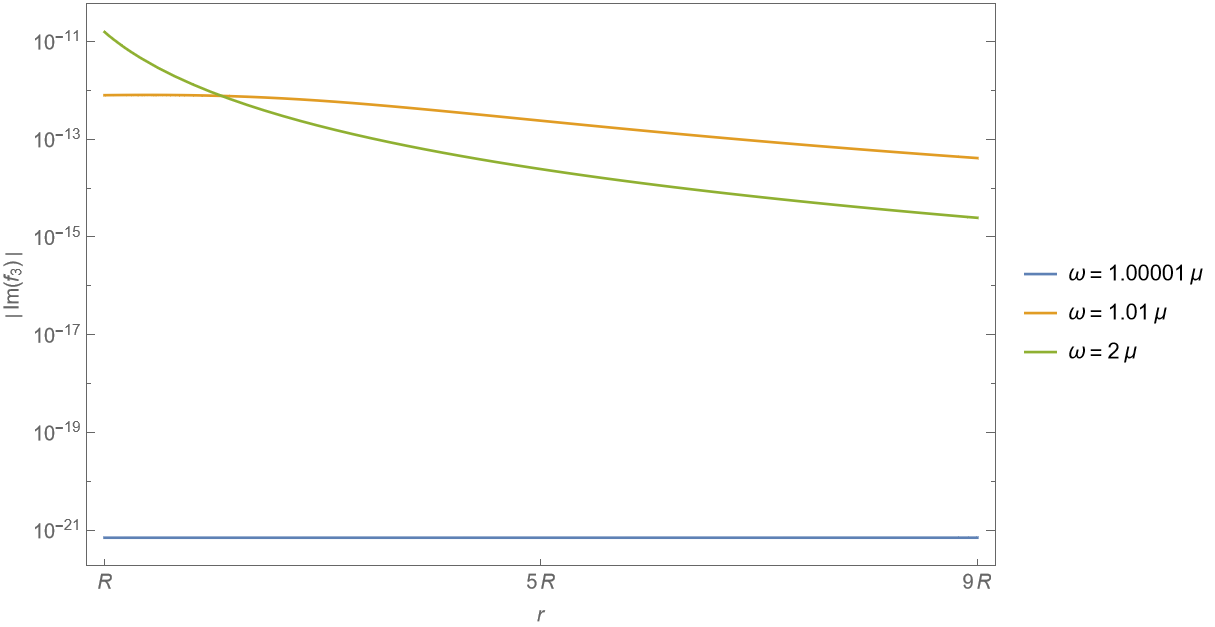}
    \caption{Absolute value of the imaginary part of $f_3$ as a function of $r$ for the different energies $\omega$ and fixed angular momentum $J = J_{ns}$. For small energies, the effect seems to be quite small, but it is possible that more terms in (\ref{F-def}) may be needed to discover the full size of the low-energy contributions.}
    \label{graph-p15-diff-vary-w-fix-J4-f-r}
\end{figure}

As we can see in figure \ref{graph-p15-diff-vary-w-fix-J4-f-r}, the angular momentum contribution is larger for higher energies. Indeed, our solution seems to take the angular momentum into account more significantly when the star is rotating quickly (though still under the “slow” rotation regime as dictated by the original paper) and the particle is a high-energy zero-spin boson.
We must remark that in the results presented in this section, the
velocity of the particle $v_p \gg v_R$, where $v_R$ is the rotation velocity of the star for $r=R$. If we have small energies, $v_p\sim v_R$, more terms will be needed in eq. (\ref{F-def}).

%
%\subsection{Comparison between the two solutions}
%

\section{Conclusions}\label{section-6}

In this work, we studied the behavior of spin-0 bosons near rotating stars. For this purpose, we considered the Klein-Gordon equation in the region external to the star with the spacetime
structure determined by the Hartle-Thorne metric (\ref{final-ht-metric}) up to first order in $j_1$, which is a very good approximation. The corrections for the metric proposed in \cite{berti2005} appear for terms of order $j_1^2$ and in general are very small. By solving the Klein-Gordon equation derived in the 
metric (\ref{final-ht-metric}), a general solution has been found, with the radial part given in terms of a confluent Heun
function. By analyzing the terms of the radial equation for
$r>R$ it is possible to neglect some of them and still have an accurate result as long as they are small. With this procedure,
we obtained a simpler solution that shows explicitly the main effect of the rotation of the star in the wave function that is given by the rotational phase defined from eq.(\ref{soldeltaR})
\begin{equation}\label{jphase}
    \delta_R=\frac{jm\omega}{2k}\ln r    \ .
\end{equation}
Then an analysis for higher powers of $j_1$ has been performed considering the Klein-Gordon equation in the Berti metric, eq.
(\ref{full-berti-metric}) and solving it by the method presented in \cite{elizalde_1987}. We presented the calculations up to $a_3$
in eq. (\ref{F-def}), and the numerical calculations have shown that for bosons with relativistic energies, the solutions converge
easily and the effect of the rotation of the stars is very small. 
Among the considered physical systems, the contribution of 
the terms depending on $j_1$ is more important for the neutron stars.
For lower values of the energies ($v_p\sim v_R$) this effect increases and more terms are needed in the solution. So, from the results we can observe that an important aspect of this effect
is that it increases as the rotation velocity of the star increases relative to the velocity of the test particle. Because of this, for stars, this effect must always be small due to the fact that $v_R$ cannot reach high values. We expect that, near black 
holes or galaxies, this effect becomes more important. This kind of behavior is similar both for pions and Higgs bosons, that
is, for light and heavy bosons.

The inclusion of mass asymmetries of the stars is straightforward, we just have to take $Q\neq 0$ in
eq. (\ref{expanded-ht-metric}) and follow the procedure presented in this work, and then, this effect may be found in the wave function.

If we think in terms of experimental results, the best way to 
study this effect is to look for observables for which the rotational phase (\ref{jphase}) is important, such as in the analysis of the
final state interactions of decaying particles or in the study of interacting particles. We must also remark that even if this kind of effect is small, an experimental verification
would be proof of the frame-dragging by rotating objects
and, consequently, of the theory of general relativity's influence on quantum systems.

\section{Acknowledgments}

We would like to thank CNPq for the financial support.

\bibliography{references}

\end{document}